
\input harvmac
\let\includefigures=\iftrue
\let\useblackboard=\iftrue
\newfam\black

\includefigures
\message{If you do not have epsf.tex (to include figures),}
\message{change the option at the top of the tex file.}
\input epsf
\def\figin{\epsfcheck\figin}\def\figins{\epsfcheck\figins}
\def\epsfcheck{\ifx\epsfbox\UnDeFiNeD
\message{(NO epsf.tex, FIGURES WILL BE IGNORED)}
\gdef\figin##1{\vskip2in}\gdef\figins##1{\hskip.5in}
instead
\else\message{(FIGURES WILL BE INCLUDED)}%
\gdef\figin##1{##1}\gdef\figins##1{##1}\fi}
\def\DefWarn#1{}
\def\figinsert{\goodbreak\midinsert}
\def\ifig#1#2#3{\DefWarn#1\xdef#1{fig.~\the\figno}
\writedef{#1\leftbracket fig.\noexpand~\the\figno}%
\figinsert\figin{\centerline{#3}}\medskip\centerline{\vbox{
\baselineskip12pt\advance\hsize by -1truein
\noindent\footnotefont{\bf Fig.~\the\figno:} #2}}
\endinsert\global\advance\figno by1}
\else
\def\ifig#1#2#3{\xdef#1{fig.~\the\figno}
\writedef{#1\leftbracket fig.\noexpand~\the\figno}%
\global\advance\figno by1} \fi

\def\id{{1 \kern-.28em {\rm l}}}

\def\K3{{\bf K3}}
\def\journal#1&#2(#3){\unskip, \sl #1\ \bf #2 \rm(19#3) }
\def\andjournal#1&#2(#3){\sl #1~\bf #2 \rm (19#3) }

\def\frac#1#2{{#1\over#2}}

\def\inbar{\,\vrule height1.5ex width.4pt depth0pt}
\def\IC{\relax\hbox{$\inbar\kern-.3em{\rm C}$}}
\def\IR{\relax{\rm I\kern-.18em R}}
\def\IP{\relax{\rm I\kern-.18em P}}

%
%

%
\catcode`\@=11
\def\slash#1{\mathord{\mathpalette\c@ncel{#1}}}
\overfullrule=0pt

\def\NN{{\cal N}}

\def\underrel#1\over#2{\mathrel{\mathop{\kern\z@#1}\limits_{#2}}}

\catcode`\@=12


%

\def \sinh{{\rm sinh}}
\def \cosh{{\rm cosh}}


\lref\BuchelTZ{
  A.~Buchel and J.~T.~Liu,
  Phys.\ Rev.\ Lett.\  {\bf 93}, 090602 (2004)
  [arXiv:hep-th/0311175].
}

\lref\KovtunWP{
  P.~Kovtun, D.~T.~Son and A.~O.~Starinets,
  JHEP {\bf 0310}, 064 (2003)
  [arXiv:hep-th/0309213].
}

\lref\HerzogGH{
  C.~P.~Herzog, A.~Karch, P.~Kovtun, C.~Kozcaz and L.~G.~Yaffe,
  JHEP {\bf 0607}, 013 (2006)
  [arXiv:hep-th/0605158].
}
\lref\HerzogSE{
  C.~P.~Herzog,
  JHEP {\bf 0609}, 032 (2006)
  [arXiv:hep-th/0605191].
}

\lref\PilchUE{
  K.~Pilch and N.~P.~Warner,
  Nucl.\ Phys.\ B {\bf 594}, 209 (2001)
  [arXiv:hep-th/0004063].
}

\lref\BuchelAH{
  A.~Buchel and J.~T.~Liu,
  JHEP {\bf 0311}, 031 (2003)
  [arXiv:hep-th/0305064].
}

\Title{\vbox{\baselineskip12pt\hbox{EFI-06-24} \hbox{}}}
{\vbox{\centerline{Friction Coefficient for Quarks in}
\bigskip
\centerline{Supergravity Duals} }}
\bigskip

\centerline{\it E. Antonyan}
\bigskip
\centerline{EFI and Department of Physics, University of
Chicago}\centerline{5640 S. Ellis Av. Chicago, IL 60637}

\smallskip

\vglue .3cm

\bigskip

\bigskip
\noindent We study quarks moving in strongly-coupled plasmas that have
supergravity duals. We compute the friction coefficient of strings dual to such
quarks for general static supergravity backgrounds near the horizon. Our results
also show that a previous conjecture on the bound has to be modified and higher
friction coefficients can be achieved.
\bigskip


\Date{November 2006}

\newsec{Introduction}

Motivated by experiments conducted at RHIC, where collisions of e.g. gold nuclei are believed to produce
strongly coupled quark-gluon plasma, studies of several properties of such plasmas have been conducted using the
AdS/CFT correspondence. Since the correspondence maps strongly coupled field theory to weakly coupled
supergravity, one could actually compute the properties of RHIC plasma if the dual of QCD was known. Even though
we do not know what the AdS/CFT dual of QCD is, some features of field theory duals are apparently universal.
One such feature is the ratio of shear viscosity to the entropy density which has been argued to have a
universal bound \KovtunWP\ for any liquid. In fact the bound was shown in \BuchelTZ\ to be saturated for gauge
theories that have supergravity duals.


Another feature that has been recently studied using the AdS/CFT correspondence is
the energy-loss of the quarks moving through strongly-coupled plasma. The study of
\HerzogGH\ was conducted for AdS$_p$ spaces with arbitrary $p$, so in particular
it was useful for studying $N=4$ supersymmetric Yang-Mills plasma which is dual to
AdS$_5$ space. It was subsequently extended to include asymptotically AdS
backgrounds \HerzogSE . Considering a particle with momentum $p$ moving in a
viscous medium subject to a driving force $f$
\eqn\drag{
    \dot{p} = -\mu p + f
} the friction (drag) coefficient $\mu$ was determined and in fact shown to be
bounded by $2\pi T$, where $T$ is the temperature of the plasma. Motivated by the
viscosity bound it was then conjectured that the bound $\mu/T\le 2\pi$ might hold
in more general theories.

In the dual picture the plasma corresponds to a black hole with temperature $T$.
To represent the quarks one has to embed an additional brane in the background of
the black hole. The quarks then correspond to strings hanging from the branes and
extending all the way to the horizon. To calculate the friction coefficient $\mu$
one can then start with a string dragged along at a constant speed that is then
let go. From \drag\ the velocity of the string will then be given by $v\propto
v_0e^{-\mu t}$, where $v_0$ is the velocity before the driving force was turned
off.

In this paper we will derive a general expression for the maximum of $\mu$ in arbitrary dual supergravity
backgrounds. We will then demonstrate that in a particular background, which is a non-extremal deformation of
the $\NN=2^*$ Pilch-Warner solution \refs{\PilchUE ,\BuchelAH}, the conjectured bound is actually violated.


\newsec{Friction coefficient}
We start by introducing the class of supergravity backgrounds that will be
considered and then proceed to computing the value of the friction coefficient
very close to the horizon.

\subsec{Supergravity backgrounds}

Let us consider static supergravity backgrounds in $D$ dimensions. Following \BuchelTZ\ we write the metric as:
\eqn\GenMetric{
    ds^2 = - \Omega^2(r)\Delta^2(r)dt^2 +
    \Omega^2(r)\left[g_{\mu\nu}(x)dx^{\mu}dx^{\nu}\right]
           + \Omega^2(r)\left[g_{rr}(r)dr^2 + g_{ij}(y)dy^i dy^j\right],
}
where $\mu,\nu = 1,\ldots,d$; $i,j=d+2,\ldots,D$ and $r$ is the radial coordinate away from the horizon, which is
stretched in $\mu$ directions. We assume that warp factors $\Delta$ and $\Omega$, as well as $g_{rr}$ depend
only on the radial coordinate. We take the horizon to be at $r=r_0$, so the non-extremality warp factor
$\Delta(r)$ vanishes at $r_0$. The warp factor $\Omega(r)$ is non-zero at the horizon, $\Omega(r_0) = \Omega_0$. One
can also show that the temperature is given by
\eqn\temp{
    T=\frac{\partial_r\Delta}{2\pi\sqrt{g_{rr}}}\Big|_{hor}
}

\subsec{Embedding a brane}

Let us embed a brane in the background \GenMetric\ at a constant radius $r_m$ as
in \HerzogGH. We will consider a string hanging from the brane down to the horizon
and will drag it along the brane to measure the friction coefficient. Thus the
brane needs to be at least 1-dimensional for our purposes. We fix all of the
$x^{\mu}$'s except for $x^1$ and absorb $g_{11}$ into the definition of $x^1$
(denoted $x$ henceforth). The string profile is then given by $x=x(r,t)$ and thus
the induced metric on the string is: \eqn\StringMetric{
    ds^2 = \left[-\Omega^2(r)\Delta^2(r) + \Omega^2(r)\dot{x}^2\right]dt^2 + 2 \Omega^2(r)\dot{x}x'dtdr
           + \left[\Omega^2(r)g_{rr}(r) + \Omega^2(r)x'^2\right]dr^2
}
where the dots and primes denote derivatives with respect to $t$ and $r$ respectively. The equation of motion
that follows from the Nambu-Goto action is then
\eqn\eom{
    \partial_r\left(\frac{\Omega^4\Delta^2x'}{\sqrt{-g}}\right) -
    \Omega^4g_{rr}\partial_t\left(\frac{\dot{x}}{\sqrt{-g}}\right) = 0,
}
where $g$ is the determinant of the induced metric \StringMetric . The simplest solution to the equation of motion
is the static straight string $x(r,t) = x_0$.

The next simplest solutions one can look for are rigid moving strings of the form
\eqn\constantv{
    x(r,t)=x_0+vt
}
However these solutions are not physical, since $-g \propto (\Delta^2-v^2)$ and
thus becomes negative close to the horizon for arbitrary $v$ and energy and
momentum are complex. Even though the time propagation is simple, the initial
conditions are inconsistent since parts of the string are moving faster than the
speed of light.

If we however allow for the string to curve, i.e. look for solutions of form
$x(r,t) = x(r) + vt$ the problem will disappear. Because the string moving on the
brane (or equivalently the quark moving in the plasma) experiences friction, such
motion at constant speed also requires an external force, e.g. an electric field
on the brane, which we assume is at hand.


Provided such an external force there are two types of solutions which are schematically represented in Fig. 1. In the first one the energy flow is toward the horizon, while in the second one the energy flows away from it. We thus adopt the postulate of \HerzogGH\ that the physical process requires us to pick the {\it purely outgoing} (with respect to the brane) boundary conditions at the horizon and so discard the second type of solutions.

\ifig\sring{For the physical solution on the left the energy flows towards the horizon, while for the unphysical solution (on the right) the energy flows away from the horizon.} {\epsfxsize3.0in\epsfbox{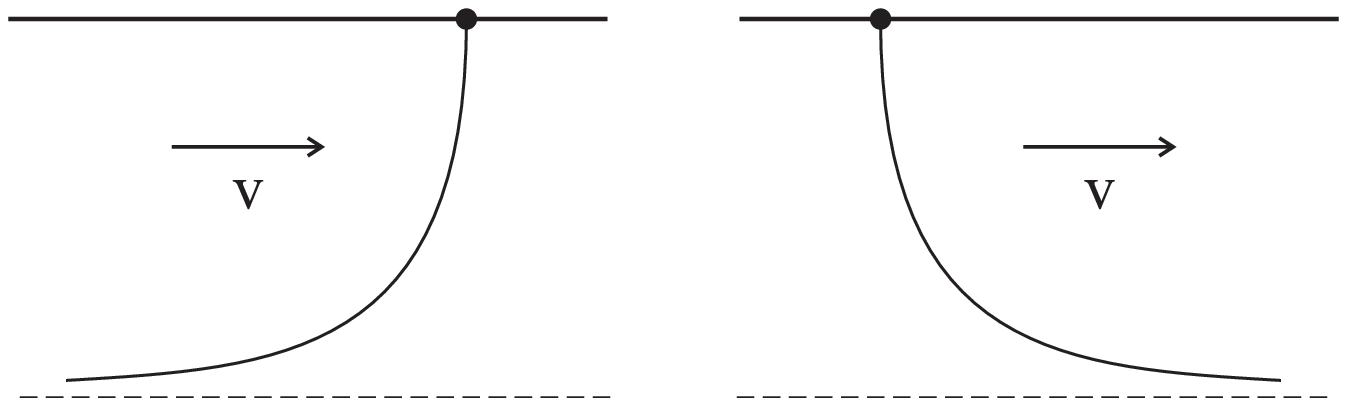}}

We now focus on late-time and thus low-velocity dynamics of the equation of motion. This corresponds to analyzing
the equation of motion in the linearized regime. Considering small fluctuations around the static straight string, i.e.
treating $\dot{x}$ and $x'$ as small and retaining only the linear terms we have:
\eqn\lineom{
    \partial_r\left(\frac{\Delta}{\sqrt{g_{rr}}}\Omega^2x'\right) = \frac{\sqrt{g_{rr}}}{\Delta}\Omega^2\ddot{x}
}

Let us rewrite the equation of motion in a covariant form.
Introducing $dw = \frac{\sqrt{g_{rr}}}{\Delta}\frac{1}{\Omega^2}dr$ we have
\eqn\coveom{
    \partial_w^2x = \Omega^4\ddot{x}
}
Close to the horizon we can solve this
\eqn\zerosln{
    x(w,t) = F(t + \Omega_0^{-2}w) + G(t - \Omega_0^{-2}w)
}
where $F(x)$ and $G(x)$ are arbitrary differentiable functions. As discussed above we follow the prescription of \HerzogGH\
and consider purely outgoing solutions, which means $G=0$.

We now restrict our attention to $e^{-\mu t}$ time dependence, corresponding to
slowing down of the string from some initial velocity, where $\mu$ is the friction
coefficient we want to find. Together with the Neumann boundary conditions at the
brane and outgoing boundary conditions near the horizon the equation of motion
becomes: \eqn\bndry{\eqalign{
        &\partial_w^2x = \mu^2\Omega^4 x \cr
        &\partial_w x|_{w_0} = 0 \cr
        &(\partial_w x + \mu\Omega^2 x)|_{\epsilon} = 0
    }
}
where $w_0$ is the location of the brane and $\epsilon$ is a point very close to the horizon (which in these coordinates
is at $w\rightarrow-\infty$). 

We have already solved the equation of motion to first non-trivial order \zerosln\ near the horizon. We now consider
an expansion $x = Ae^{-\mu t}e^{-\mu\Omega_0^2w}(1+\delta x + O((\delta x)^2))$, where $\delta x \ll 1$ (note that this
automatically takes care of the outgoing boundary conditions). Also expand $\Omega(w)$ near the horizon as $\Omega(w) =
\Omega_0 + \Omega_1(w) + \ldots$, where $\Omega_1(w) \ll \Omega_0$ near the horizon. One can then solve the differential
equation in this approximation to get
\eqn\solution{
    \delta x = 4\mu^2\Omega_0^3\int_{-\infty}^{w}dw'e^{2\mu\Omega_0^2w}\int_{-\infty}^{w'}dw''e^{-2\mu\Omega_0^2w''}\Omega_1(w'')
} If we want to satisfy the Neumann boundary conditions to this order we then get
a condition on the friction coefficient $\mu$: \eqn\musln{
    4\mu\Omega_0e^{2\mu\Omega_0^2w_0}\int_{-\infty}^{w_0}dw'e^{-2\mu\Omega_0^2w'}\Omega_1(w') -
    4\mu^2\Omega_0^3\int_{-\infty}^{w_0}dw'e^{2\mu\Omega_0^2w}\int_{-\infty}^{w'}dw''e^{-2\mu\Omega_0^2w''}\Omega_1(w'') = 1
}
Given the metric one can expand $\Omega(w)$ and plug into the above equation to solve for $\mu(w_0)$.

For a class of metrics found most commonly in the literature this can actually be solved when the brane is very close to
the horizon $w_0\rightarrow -\infty$. In particular when $\Omega(w)$ has the following form near the horizon:
$\Omega(w) = \Omega_0 + C e^{\alpha w} + \ldots$, where $C$ and $\alpha>0$ are constants. In the original coordinates
$r$ this is equivalent to $\Omega(r)$ having a polynomial expansion in $r$, which is what one generically finds.
Solving for $\mu$ as the brane location is taken to be close to the horizon we get the limiting value
\eqn\mueqn{
    \mu = \frac{\partial_w\Omega}{2\Omega_0^2(\Omega-\Omega_0)}\Big|_{hor}
}
or in the original $r$ coordinates
\eqn\mureqn{
    \mu = \frac{\Delta}{2\sqrt{g_{rr}}}\frac{\partial_r\Omega}{\Omega-\Omega_0}\Big|_{hor}
}
One can then hope that this expression gives the bound for $\mu$, since $\mu\rightarrow 0$ as the probe brane location
is moved to be very far from the horizon. That is exactly what happened for the solutions considered in \refs{\HerzogGH ,\HerzogSE} where the
numerical investigations gave a monotonic curve for $\mu$ as a function of the brane location.

Comparing this with the expression for the temperature we see that the bound $\mu=2\pi T$ suggested in \HerzogGH\ will be
satisfied iff
\eqn\muTcond{
    \Delta|_{hor} \propto (\Omega - \Omega_0)^{\frac{1}{2}}|_{hor}
}
This relation is indeed satisfied for many supergravity backgrounds, in particular those considered in \refs{\HerzogGH ,\HerzogSE}, as well as
e.g. the background of a stack of parallel branes of arbitrary dimensionality. We now however consider an example where it is actually violated
and $\mu = 4\pi T$ close to the horizon. 

The background we consider is the non-extremal deformation of the Pilch-Warner solution \PilchUE, considered in \BuchelAH. This solution realizes the supergravity dual to $\NN=4$, $SU(N)$ gauge theory that is softly broken to $\NN=2$.

As shown in \BuchelAH\ the 5-d Einstein frame finite temperature metric is given by
\eqn\pwmetric{
    ds_5^2 = e^{2A}(-e^{2B}dt^2 + d\vec{x}^2) + dr^2
}
The horizon is taken to be at $r = 0$ and the metric coefficients have the following near horizon
expansions:
\eqn\expansion{\eqalign{
        &e^A = e^{\alpha}\left(1+\sum_{i=1}^{\infty}a_ir^{2i}\right) \cr
        &e^B = \delta r\left(1+\sum_{i=1}^{\infty}b_ir^{2i}\right)
    }
}
where $\alpha,\delta,a_i,b_i$ are constants. Ultimately the solution is
parameterized by a three parameter family $\{\alpha,\rho_0>0,\chi_0\}$ and in fact
for $i=1$ \BuchelAH\ determined $a_1$ and $b_1$ to be
\eqn\ab{\eqalign{
        &\delta^{-2}a_1 = \frac{1}{12}\rho_0^{-4} + \frac{1}{6}\rho_0^2\cosh(2\chi_0)-\frac{1}{48}\rho_0^8\sinh^2(2\chi_0) \cr
        &b_1 = -\frac{4}{3}a_1
    }
}
The functions $\Delta$ and $\Omega$ of \GenMetric\ are then equal to
\eqn\deltas{\eqalign{
        &\Delta = e^{B} = \delta r(1+b_1r^2+\ldots)\cr
        &\Omega = e^{A} = e^{\alpha}(1+a_1r^2+\ldots)
    }
}
For generic values of $\rho_0$ and $\chi_0$, $a_1$ and $b_1$ are non-zero and thus \muTcond\ is satisfied, however as can be seen from Fig. 2 for special values of $\rho_0$ and $\chi_0$, $a_1=b_1=0$ and thus (assuming that $a_2$ is non-zero) we actually have $\mu=4\pi T$.

\ifig\sring{Plot of the coefficient $\delta^{-2}a_1$ for $\chi_0=1$.} {\epsfxsize3.0in\epsfbox{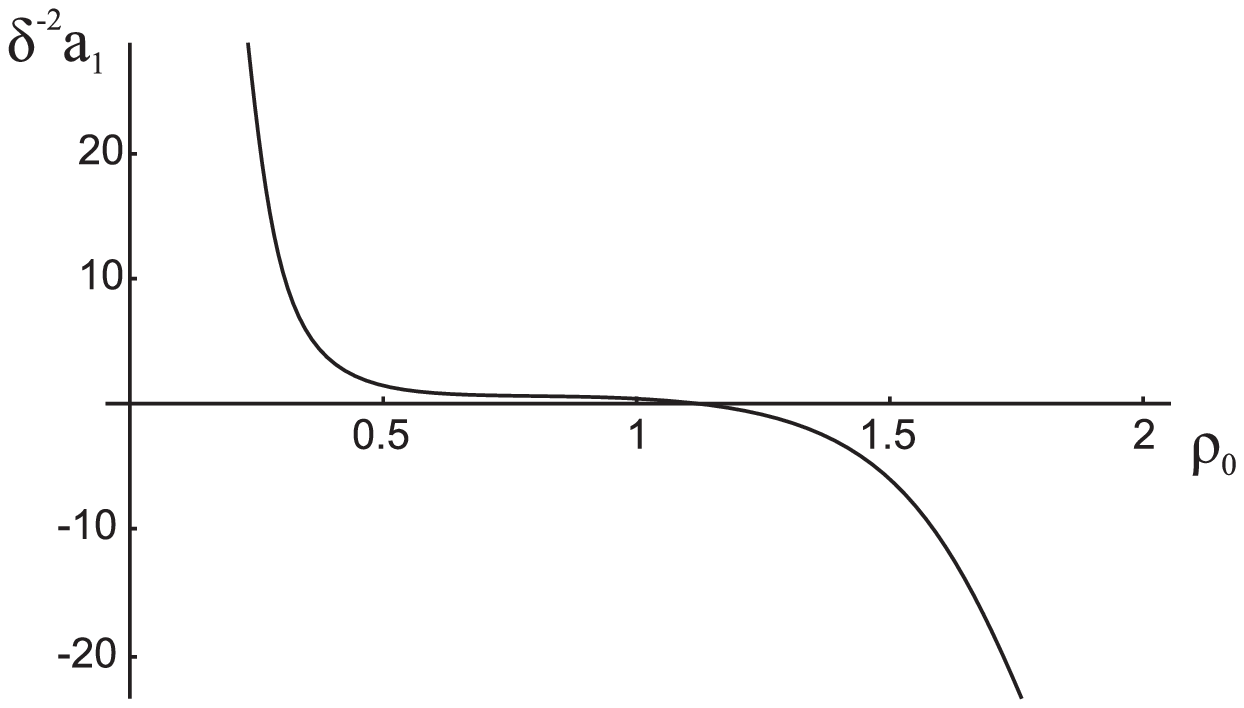}}


\newsec{Discussion}
In this paper we derived the expression for the friction coefficient of quarks
moving in strongly-coupled plasmas that have a supergravity dual. It was
conjectured in \HerzogGH\ and consequently in \HerzogSE\ that the friction
coefficient $\mu$ has a universal maximum equal to $2\pi T$. However as we have
shown above there is a solution for which $\mu$ can be as high as $4\pi T$. It is
not clear whether there are any solutions that will have even bigger friction
coefficients, however no general argument seems to put a restriction on that.

A related issue is whether or not $\mu(w_0)$ is a monotonic function of the location of probe brane $w_0$. In the examples considered in the literature \refs{\HerzogGH ,\HerzogSE} one finds that it is, but that result requires explicit numerical computations. We leave the investigation of this question to future research.

\bigskip\medskip\noindent
{\bf Acknowledgements:} We thank J. Harvey, O. Lunin and R. Wald for discussions. This work was supported in part by NSF Grant
No. PHY-0204608.


\listrefs
\end